# Plasmonic structure integrated single-photon detector configurations to improve absorptance and polarization contrast

Mária Csete[*], Gábor Szekeres, András Szenes, Anikó Szalai and Gábor Szabó

*Department of Optics and Quantum Electronics, University of Szeged, H-6720 Szeged, Dóm tér 9, Hungary*
[*]*mcsete@physx.u-szeged.hu*

**Abstract:** Configurations capable of maximizing both absorptance and polarization contrast were determined for 1550 nm polarized light illumination of different plasmonic structure integrated superconducting nanowire single-photon detectors (SNSPDs) consisting of $p$=264 nm and $P$=792 nm periodic niobium-nitride (NbN) patterns on silica substrate. Global NbN absorptance maxima appear in case of p/s-polarized light illumination in S/P-orientation ($\gamma$=90°/0° azimuthal angle) and the highest polarization contrast is attained in S-orientation of all devices. Common nanophotonical origin of absorptance enhancement is collective resonance on nano-cavity-gratings with different profiles, which is promoted by coupling between localized modes in quarter wavelength MIM nano-cavities and laterally synchronized Brewster-Zenneck-type surface waves in integrated SNSPDs possessing a three-quarter-wavelength-scaled periodicity. The spectral sensitivity and dispersion characteristics reveal that device design specific optimal configurations exist.

©2014 Optical Society of America

## 1. Introduction

Superconducting nanowire single-photon detectors (SNSPDs) are important devices for telecommunication, single-photon counting and quantum information processing application areas [1-3]. To improve detection efficiency the effective absorption cross-section of these devices has to be maximized, while to maintain short reset time the kinetic inductance has to be minimized. These two opposite demands are challenging for SNSPD designers, and recent tradeoff in SNSPD development is ensuring the highest possible absorptance via the shortest possible superconducting wire-patterns [4, 5].

To achieve the highest possible detection efficiency, different absorbing materials, light in-coupling methodologies and device designs have been inspected. The prevalent structure is a strongly sub-wavelength (~200 nm) periodic meandered pattern of 4-5 nm thick, 10-100 nm wide bare superconducting stripes on a substrate. Conventional absorbing material in commercially available SNSPD devices is niobium-nitride (NbN), while the substrate is dominantly sapphire due to its advantageous properties [1, 6]. Novel superconducting materials acting as absorbing elements were investigated as well, and ~93 % absorptance was attained via 200 nm pitch WSi wires embedded into a silicon based optical stack [7].

The efforts to find the optimal illumination methodology resulted in $\gtrsim 90$ % absorptance via sub-wavelength patterns in optical fiber coupled [7] and in waveguide coupled devices as well [8]. Moreover, different integrated structures were successfully applied to improve absorptance [9-15]. Theoretical studies predicted ~43 % and ~68 % absorptance in devices consisting of sub-wavelength NbN patterns integrated with a dielectric cavity (DC-SNSPD) [11] and with an optical-cavity closed by a gold reflector (OC-SNSPD) [10], while 54 % absorptance was observed experimentally in OC-SNSPDs at perpendicular incidence [9].

Complex integrated plasmonic structures were shown to result in significantly enhanced infrared light absorption with pronounced polarization, illumination direction and wavelength sensitivity [12-15]. Our latest results prove that 93.4 % / 67.7 % absorptance is attainable in nano-cavity-arrays with 220 nm / 660 nm periodicity at large tilting, where the absorptance is almost wavelength independent. Moreover, integration of wavelength-scaled nano-cavity-deflector-arrays comprising gold segments penetrating into the sapphire substrate results in 93.3 % / 81.6 % absorptance at smaller tilting. The absorptance is considerably large throughout a ~100 nm wavelength interval surrounding the center of inverted plasmonic band-gaps appearing in three-quarter-wavelength patterns [15]. However, fabrication of sapphire substrates is extremely challenging, as a consequence initialization of analogous nanophotonical phenomena in absorbing patterns on different substrates is necessary.

The motivation of our research was to develop plasmonic structure integrated SNSPD devices, and to determine device configurations with potential to realize single plasmon detection [16] and to read-out encoded quantum information with good fidelity [17].

The purposes of present work were to inspect the illumination direction and wavelength dependence of the attainable absorptance and near-field distribution of SNSPDs integrated with versatile plasmonic structures. Specific objectives were to understand the nanophotonical origin of absorptance modulation and to determine those integrated device configurations, which are optimal for polarization selective read-out. The selection of silica substrate for integrated SNSPD design ensures the possibility of future experimental realization.

**2. Methods**

Four different types of plasmonic structure integrated SNSPD devices were inspected theoretically, and for each design a sub-wavelength periodicity as well as a periodicity commensurate with $\lambda_{SPP}$ wavelength of surface plasmon polaritons (SPPs) excitable at silica substrate and gold interface was considered [10, 11, 14, 15]. The small pitch devices consist of meandered NbN patterns with $p \sim 1/4 * \lambda_{SPP, 1550\,nm}$ = 264 nm periodicity, while in large pitch devices the absorbing patterns have a periodicity of $P = 3/4 * \lambda_{SPP, 1550\,nm}$ = 792 nm. The 4 nm thick and 100 nm wide NbN wires are positioned at the entrance of MIM nano-cavities having a sub-wavelength width of 100 nm. The 220 nm nano-cavity-length corresponds to quarter-wavelength of squeezed MIM modes that are resonant at 1550 nm in single 100 nm wide HSQ (hydrogen-silsesquioxane) filled channels embedded into gold medium. All nano-cavities are closed by 60 nm thick horizontal gold layer and by arrays of vertical gold segments with different geometry composing different kinds of MIM nano-cavity-gratings (Fig. 1).

In nano-cavity-array integrated NCAI-SNSPDs each NbN stripe is surrounded by 164 nm and 692 nm wide vertical gold segments composing a $p$ = 264 nm and $P$ = 792 nm periodic MIM nano-cavity-grating, respectively. These integrated devices possess a symmetric surface profile with respect to MIM cavity centers (Fig. 1aa, ba) [14, 15].

In nano-cavity-deflector-array integrated NCDAI-SNSPDs 100 nm wide and 220 nm long gold segments are inserted into the silica substrate at 792 nm distances at the anterior side of each third and each NbN loaded nano-cavity in $p$- and $P$-pitch design, respectively. Thus a secondary deflector grating with a periodicity of 792 nm is created, however with an asymmetric profile (Fig. 1ab, bb) [15]. The deflectors were designed according to the literature about efficient plasmonic mirrors, namely their $\sim \lambda_{SPP, 1550\,nm}/10$ width is capable of ensuring high reflection efficiency. Moreover, the 220 nm deflector depth is commensurate with $0.2 * \delta_{SPP, 1550\,nm}$, where $\delta_{SPP, 1550\,nm}$ is the transversal decay length of the SPP's in silica, as a result this length still makes possible good SPP reflection [15, 18].

In nano-cavity-double-deflector-array integrated NCDDAI-SNSPDs 100 nm wide and 220 nm long deflectors are inserted into the silica substrate at the middle of each 264 nm wide vertical gold segment in $p$-pitch pattern, while analogous deflectors neighbor each nano-cavity both at their anterior and exterior sides in $P$-pitch pattern (Fig. 1ac, bc).

In nano-cavity-trench-array integrated NCTAI-SNSPDs 220 nm deep trenches having a width of 100 nm and 492 nm are embedded into the in-plane vertical gold segments in $p$- and $P$-pitch design, respectively (Fig. 1ad, bd). Although, the 492 nm width of the trenches in $P$-pitch NCTAI-SNSPD is larger than the $\sim \lambda_{SPP, 1550\,nm}/10$ width of deflectors, they can still act as efficient plasmonic mirrors, due to the oscillatory behavior of reflectance in case of trenches as a function of their width [18]. Moreover, insertion of wider trenches makes it possible to reduce the amount of gold, which causes competitive absorption. NCTAI-SNSPDs can be considered as special nano-cavity-arrays with 32 nm and 100 nm wide cavity walls in the $p$- and $P$-pitch design, respectively.

Theoretical studies were performed to determine the optimal illumination directions for each SNSPD device designs, using finite element method we have previously developed based on the Radio Frequency module of COMSOL Multiphysics software package (COMSOL AB) [10, 11, 14, 15]. Three dimensional models were applied to study the polar ($\varphi$) and azimuthal ($\gamma$) angle dependence of different plasmonic structure integrated SNSPDs optical response (Fig. 1c, d, Fig, 2, 3). The polar angle dependent polarization contrast of each device designs was also determined in P- and S-orientation (Fig. 4).

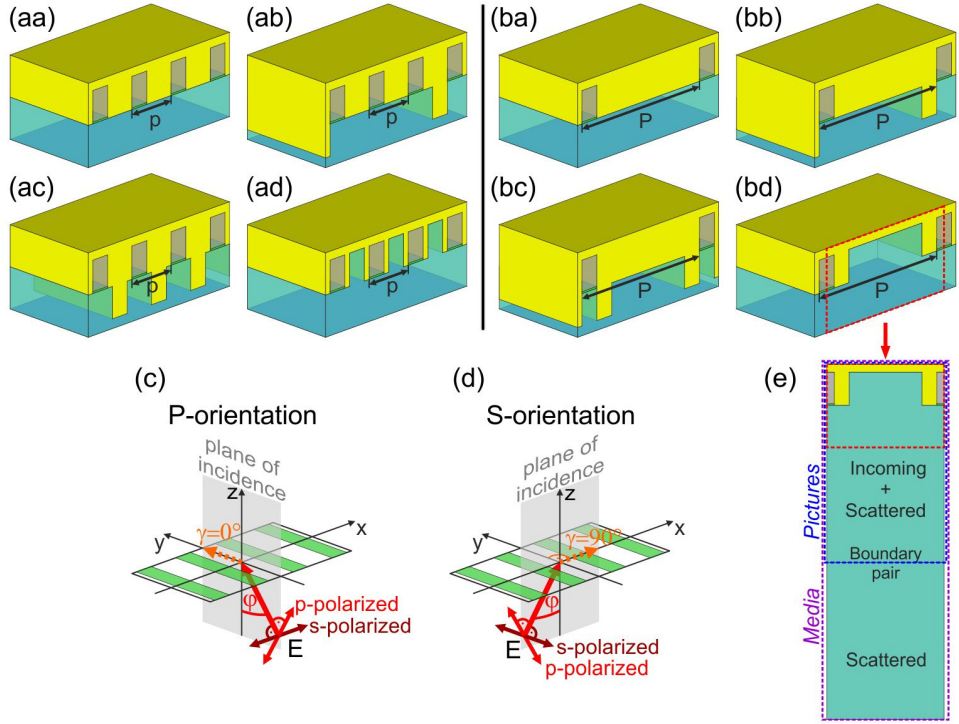

Fig. 1. SNSPD devices on silica substrate consisting of (aa-ad) $p$ = 264 nm periodic and (ba-bd) 792 nm periodic pattern of 4 nm thick and 100 nm wide NbN stripes integrated with (aa, ba) quarter-plasmonic-wavelength nano-cavity-array (NCAI-SNSPD), (ab, bb) nano-cavity-deflector-array (NCDAI-SNSPD), (ac, bc) nano-cavity-double-deflector-array (NCDDAI-SNSPD), and (ad, bd) nano-cavity-trench-array (NCTAI-SNSPD). (c, d) The methodology of polar ($\varphi$) and azimuthal ($\gamma$) angle variation during p- and s-polarized infrared ($\lambda$ = 1550 nm) light illumination of plasmonic structure integrated SNSPDs from silica substrate side. Specific presented illumination configurations are: (c) P-orientation and (d) S-orientation. (e) Plane cross-section of a single unit cell used for near-field study, above/below boundary pair in pictures/Media.

The spectral sensitivity of plasmonic structure integrated SNSPD devices was analyzed by interrogating the optical response in those configurations, which result in NbN absorptance maxima with potential interest to practical applications (Fig. 5, 6). The polarization contrast achievable at specific tilting corresponding to NbN maxima in P- and S-orientation of different designs was also determined in a 500 nm wide band around 1550 nm (Fig. 7). The index of refraction of dielectric materials (silica, $NbNO_x$ and HSQ) was specified via Cauchy formulae, while dielectric constants for both metallic materials (NbN, Au) were loaded from tabulated datasets. The time-averaged **E**-field distribution was studied along with the power-flow at plane cross-sections taken perpendicular to the single unit cells of the $p$- and $P$-pitch patterns (Fig. 1e, Fig. 8, Fig. 9). The **E**-field time evolution was inspected as well to characterize all localized and propagating modes supported by the integrated devices and is provided in multimedia files (Fig. 1e, Fig. 8, 9 with Media 1-20).

### 3. Results

*3.1 Illumination direction dependent NbN absorptance and polarization contrast*

Inspection of illumination direction's effect on the NbN absorptance at 1550 nm wavelength proved that device orientations corresponding to **E**-field oscillation direction perpendicular to integrated vertical gold segments are preferred (Fig. 2a, b and Fig. 3a, b with insets). Namely, in all plasmonic structure integrated NCAI-, NCDAI-, NCDDAI- and NCTAI-SNSPDs p-polarized light is detected with higher efficiency in S-orientation (Fig. 2a, b), while in case of s-polarized light illumination P-orientation results in larger absorptance (Fig. 3a, b). These specific illumination directions are referred as p-onto-S and s-onto-P configurations throughout the paper.

*3.1.1 Illumination direction dependent NbN absorptance: p-polarization*

The p-polarized light absorptance in NbN weakly depends on the illumination direction in all *p*-pitch devices. The absorptance exhibits a global maximum at large tilting at all azimuthal orientations in NCAI-SNSPD and throughout a well defined azimuthal-polar angle interval in NCDAI-SNSPD, while it increases monotonously either by increasing the azimuthal angle, or by decreasing the polar angle in NCDDAI- and NCTAI-SNSPD (Fig. 2a with insets).

The largest 95.3 % global NbN absorptance maximum appears in S-orientation of *p*-pitch NCAI-SNSPD at 60.0° tilting, which corresponds to the plasmonic Brewster angle (PBA) of the sub-wavelength nano-cavity-grating (Fig. 2aa) [15, 19-21]. In *p*-pitch NCDAI-SNSPD a smaller 92.7 % global NbN absorptance maximum appears at 19.9° polar angle corresponding to tilting, which results in global NbN absorptance maximum in counterpart *P*-pitch design as well. This indicates that well-defined absorptance modulation is caused by the *P*-periodic deflector pattern (Figs. 2ab - to - 2bb). This maximum is overridden by the 93.3 % global NbN absorptance maximum observable at 18.0° tilting in *p*-pitch NCDDAI-SNSPD, however this extremum appears at a slightly smaller polar angle, than the tilting resulting in a global maximum in counterpart *P*-pitch design (Figs. 2ac - to - 2bc).

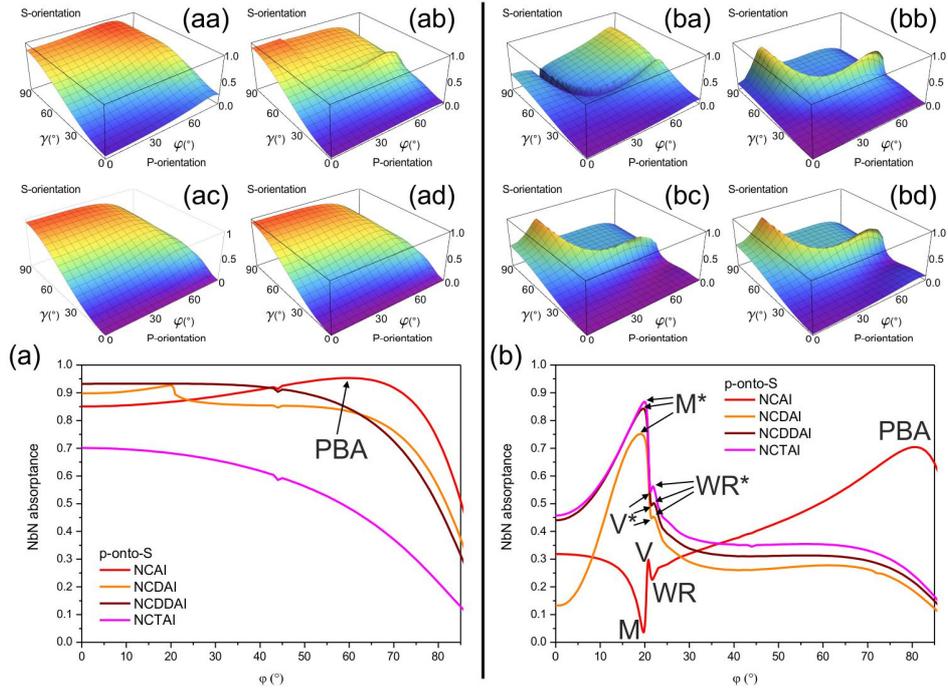

Fig. 2. Illumination direction dependence of NbN absorptance in *p*-pitch and *P*-pitch integrated devices, (aa and ba) NCAI-SNSPD, (ab and bb) NCDAI-SNSPD, (ac and bc) NCDDAI-SNSPD, and (ad and bd) NCTAI-SNSPD illuminated by 1550 nm p-polarized light. Comparison of polar angle dependent NbN absorptances in S-orientation of (a) $p$ = 264 nm, (b) $P$ = 792 nm periodic integrated SNSPDs illuminated by p-polarized light.

In NCDDAI-SNSPD caused by the absence of *P*-periodic pattern the absorption does not exhibit a local modulation, it decreases monotonously by tilting with a slightly larger rate than in *p*-pitch NCDAI-SNSPD (Figs. 2ac - to 2ab). Significantly smaller global NbN absorptance maximum of 70.1 % is reached at perpendicular incidence in NCTAI-SNSPD, however fingerprint of the modulation in counterpart *P*-pitch design is not observable (Figs. 2ad - to - 2bd), on the contrary, the NbN absorptance monotonously decreases by tilting.

Comparison of NbN absorptances achievable in *p*-pitch SNSPDs shows that at perpendicular incidence the absorptance is enhanced in presence of both single and double deflectors with respect to that in NCAI-SNSPD (Fig. 2ab, ac - to - aa), while presence of embedded trenches is not advantageous neither at perpendicular nor at oblique incidence (Fig. 2ad). Although, plasmonic Brewster angle related maxima do not appear in NCDAI- and NCDDAI-SNSPDs, and the attained NbN absorptance maxima are slightly smaller, these maxima appear at smaller polar angles, which is advantageous in practical applications (Fig. 2ab, ac).

In all *P*-pitch integrated devices sudden modification is observable on the p-polarized light absorptance in similar azimuthal-polar angle intervals (Fig. 2b insets). In NCAI-SNSPD global NbN absorptance minima are observable (Fig. 2ba), while global maxima appear in both deflector and embedded trench integrated devices, indicating that design specific orientations exist, which are advantageous for NbN absorptance maximization (Fig. 2bb-bd).

In S-orientation of *P*-pitch NCAI-SNSPD the NbN absorptance is 31.8 % at perpendicular incidence, and increases almost monotonously through a global maximum appearing at large tilting except a narrow polar angle region, where significant absorptance modulation is observable (Fig. 2ba). The NbN absorptance reaches 3.6 % global minimum at 19.7° tilting (M point), which is followed by a 29.6 % local maximum at 20.8° polar angle (V point) and a 22.6 % local minimum at 21.7° tilting (WR point). Considerably larger NbN absorptance of 70.3 % is attained at 80.0° tilting corresponding to plasmonic Brewster-angle in *P*-pitch nano-cavity-grating [15, 19-21]. Although, the attained absorptance is smaller than in counterpart *p*-pitch design, it is significantly larger than the value extrapolated by taking into account three-times smaller fill-factor.

In S-orientation of *P*-pitch NCDAI/NCDDAI/NCTAI-SNSPDs the extrema observable at small tilting are inverted compared to those in NCAI-SNSPD (Fig. 2b with insets). In NCDAI/NCDDAI/NCTAI-SNSPD 75.0 % / 84.3 % / 86.7 % global NbN absorptance maximum appears at 19.4° / 19.7° / 20.0° polar angle (M* point) (Fig. 2bb, bc, bd), and all these maxima override the absorptance maximum observable at PBA in NCAI-SNSPD. The global maximum is followed by 45.1 % / 49.3 % / 54.4 % local minimum at 21.5° / 21.5° / 21.3° polar angle (V* point) and by 45.5 % / 50.2 % / 56.2 % local maximum at 22.0° / 22.0° / 21.8° tilting (WR* point).

Comparison of NbN absorptances achievable at perpendicular incidence shows that in *P*-pitch NCDAI-SNSPD the absorptance is considerably smaller than in NCAI-SNSPD, in contrast to *p*-pitch patterns. In NCDDAI- and NCTAI-SNSPD larger NbN absorptance is achieved from perpendicular incidence throughout ~30.0° polar angle. In NCTAI-SNSPD the highest NbN absorptance peak is accompanied by intermediate FWHM, in NCDDAI-SNSPD an intermediate absorptance maximum appears with the largest FWHM, while the smallest NbN absorptance is achieved in the narrowest angle interval in NCDAI-SNSPD (Fig. 2b).

*3.1.2 Illumination direction dependent NbN absorptance: s-polarization*

The absorptance of s-polarized light in NbN segments weakly depends on the illumination direction in all *p*-pitch devices and monotonously increases by decreasing either the azimuthal angle, or the polar angle, except in *p*-pitch NCDDAI-SNSPD and NCTAI-SNSPD, which exhibit shallow global maxima at transitional tilting (Fig. 3a with insets).

The 83.7 % and 88.6 % NbN absorptance maxima are reached at perpendicular incidence of s-polarized light onto *p*-pitch NCAI-SNSPD and NCDAI-SNSPD in P-orientation (Fig. 3aa, ab). The absorptance monotonously decreases with similar rate in NCAI- and NCDAI-SNSPD by increasing the tilting.

Both absorptances are overridden by the NbN absorptance in *p*-pitch NCDDAI-SNSPD, which exhibits 93.3 % global NbN absorptance maximum at transitional 15.0° tilting (Fig. 3ac). The achievable absorptance is the smallest in NCTAI-SNSPD (Fig. 3ad) at any illumination direction, an the global maximum of 70.2 % appears at 14.0° tilting.

Comparison of the achievable NbN absorptances shows enhancement in presence of both single and double deflectors with respect to NCAI-SNSPD, while embedded trenches cause decrease throughout the entire polar angle interval also in case of s-polarized illumination. There is a significant difference in NbN absorptance polar angle dependencies in p-onto-S and s-onto-P configurations of *p*-pitch NCAI- and NCDAI-SNSPD. This indicates that unique nanophotonical phenomena are at play in these devices in p-onto-S configuration. In contrast, the global maxima equal and almost coincide in s-onto-P and p-onto-S configurations of NCDDAI- and NCTAI-SNSPDs. The collective resonances are optimized with the highest/smallest efficiency in presence of double deflectors/embedded trenches almost throughout the entire polar angle interval in both configurations.

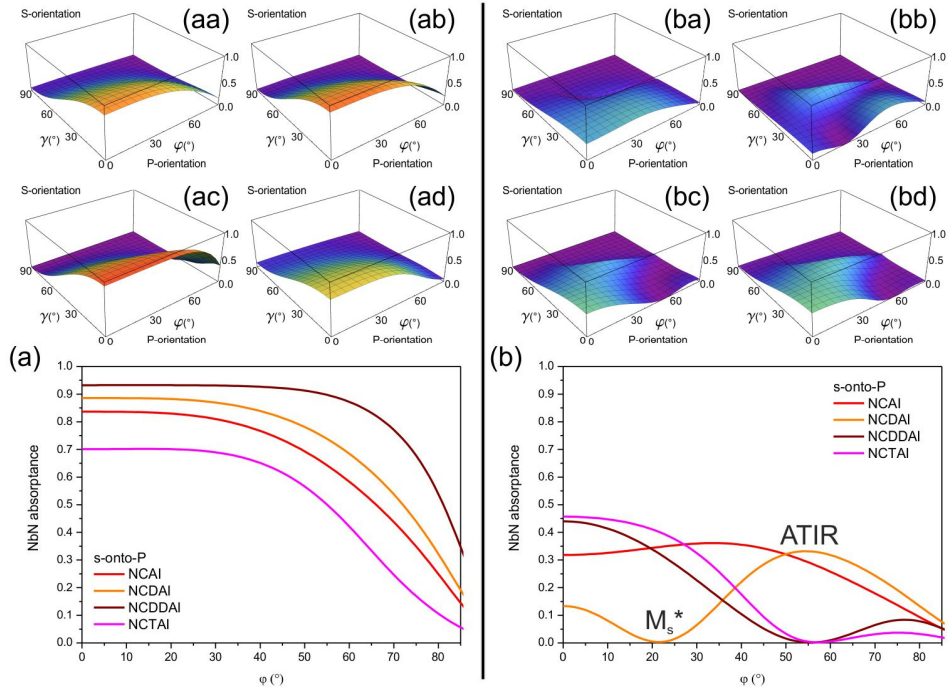

Fig. 3. Illumination direction dependence of NbN absorptance in *p*-pitch and *P*-pitch integrated devices, (aa and ba) NCAI-SNSPD, (ab and bb) NCDAI-SNSPD, (ac and bc) NCDDAI-SNSPD, and (ad and bd) NCTAI-SNSPD illuminated by 1550 nm s-polarized light. Comparison of polar angle dependent NbN absorptances in P-orientation of (a) $p = 264$ nm, (b) $P = 792$ nm periodic integrated SNSPDs illuminated by s-polarized light.

In *P*-pitch integrated SNSPD devices a smaller modulation is observable on the s-polarized light absorptance in analogous polar-azimuthal angle intervals, where sudden absorptance modifications occur in case of p-polarized illumination (Fig. 3b to Fig. 2b with insets). Namely, in NCAI-SNSPD local minima are observable, while local maxima appear in similar azimuthal-polar angle intervals in both deflector integrated devices as well as in presence of embedded trenches. This indicates that device design specific orientations exist that are advantageous for absorptance maximization in case of s-polarized light illumination.

NbN absorptance modulation phenomena become more significant in P-orientation of all *P*-pitch devices, however the polar angle interval corresponding to maximal absorption strongly depends on the integrated structure (Fig. 3b). In *P*-pitch detectors the global NbN absorptance maxima are considerably smaller than in counterpart *p*-pitch devices. In case of NCAI-SNSPD the NbN absorptance reaches 36.1 % global maximum at 34.0° (Fig. 3ba). NCDAI-SNSPD shows the smallest absorptance at perpendicular incidence among *P*-pitch devices. In addition to this a 0.4 % global NbN absorptance minimum appears at 21.0°, than a 33.2 % global maximum is achieved at 54.0° tilting, which is inside the polar angle region corresponding to ATIR phenomenon (Fig. 3bb). The appearance of a global minimum in s-onto-P configuration close to tilting corresponding to global maximum in p-onto-S configuration indicates that in latter case polarization specific phenomena are at play. In NCDDAI-SNSPDs the 44.0 % global NbN absorptance maximum appears at perpendicular incidence in P-orientation, which overrides both maxima attainable in *P*-pitch NCAI- and NCDAI-SNSPD devices. This maximum is followed by a 0.1 % global minimum at 56.0° polar angle (Fig. 3bc). The 45.7 % global NbN maximum achieved at perpendicular incidence onto NCTAI-SNSPD is the highest among absorptances reached via s-polarized illumination of P-pitch devices. The NbN absorptance exhibits a course similar to that in NCDDAI-SNSPD, however it is larger throughout the 0.2 % global minimum at 57.0° (Fig. 3bd).

The almost coincident global minima in NCDDAI- and NCTAI-SNSPDs appear close to the global maximum in s-onto-P configuration of NCDAI-SNSPD, which indicates that asymmetric deflector profiles can be advantageous in specific configurations. Inevitable advantage of double deflectors and embedded trenches is that the largest NbN absorptance is achievable from perpendicular incidence through 19.0° and 27.0° polar angles in their presence, respectively.

*3.1.3 Illumination direction dependent polarization contrast*

The polar angle dependent polarization contrast was determined both for S- and P-orientation of all plasmonic structure integrated SNSPDs, by dividing the p/s-polarized NbN absorptance by the s/p-polarized one, respectively (Fig. 4). The polarization contrast is implicitly the same at perpendicular incidence, than exhibits a polar angle dependency, which fundamentally differs in the two device orientations, and shows a characteristics influenced strongly by the integrated SNSPD designs.

**Table 1. Polarization contrast**

| | p = 264 nm | | | | P = 792 nm | | | |
|---|---|---|---|---|---|---|---|---|
| | p-to-s in S-orientation | | s-to-p in P-orientation | | p-to-s in S-orientation | | s-to-p in P-orientation | |
| Design | $\varphi$ (°) | Contrast | $\varphi$ (°) | Contrast | $\varphi$ (°) | Contrast | $\varphi$ (°) | Contrast |
| NCAI | 80.0 | 63.2 | 0.0 | 13.6 | 80.0 | 192.8 | 0.0 | 15.8 |
| NCDAI | 80.0 | 110.1 | 0.0 | 29.6 | 80.0 | 1196.2 | 0.0 | 54.1 |
| NCDDAI | 80.0 | 8916.3 | 0.0 | 2902.9 | 80.0 | 172632.9 | 0.0 | 47839.7 |
| NCTAI | 80.0 | 11.4 | 0.0 | 7.1 | 20.5 | 53.2 | 0.0 | 18.5 |

In *p*-pitch integrated devices by increasing the polar angle the polarization contrast monotonously increases in S-orientation, while in P-orientation monotonous decrease is observable (Fig. 4a). The polarization contrast achievable at perpendicular incidence is commensurate of *p*-pitch integrated NCAI-, NCDAI-, and NCTAI-SNSPDs. The presence of single deflector in NCDAI-SNSPD does not modify the course of polarization contrast significantly, namely, single deflector results in higher contrast already at perpendicular incidence and in a slightly steeper curve throughout the entire polar angle interval with respect to NCAI-SNSPD. A slight polarization contrast modulation appears near 20.0° polar angle corresponding to p-polarized absorptance modulation. In NCDDAI-SNSPD ~2.9·10$^3$ contrast is reached already at perpendicular incidence, than the p-to-s contrast increases with a larger rate. These results indicate that double deflectors have significant polarization contrast improving role.

Presence of embedded trenches is not advantageous, on the contrary, in NCTAI-SNSPD the contrast is smaller already at perpendicular incidence, and increases through the entire polar angle interval with a smaller rate than in NCAI-SNSPD. Polarization contrast maximum of $6.3 \cdot 10^1$, $1.1 \cdot 10^2$ and $1.1 \cdot 10^1$ is attained at ~80.0° tilting in NCAI-, NCDAI-, and NCTAI-SNSPD, respectively, while a significantly larger $8.9 \cdot 10^5$ contrast is achieved in NCDDAI-SNSPD.

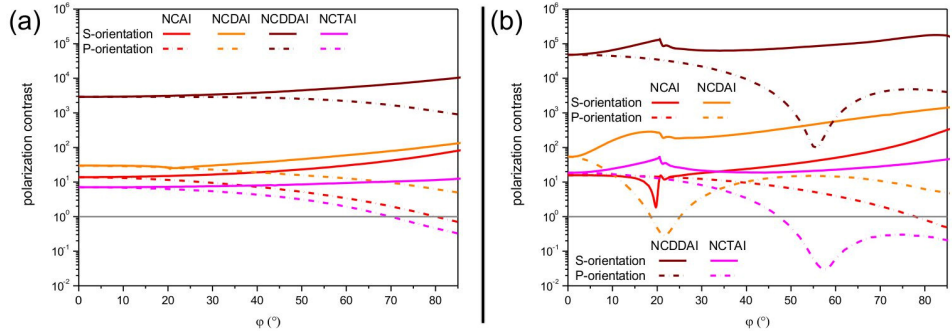

Fig. 4. Comparison of polar angle dependent polarization contrast achievable in P- and S-orientation of (a) $p$ = 264 nm, (b) $P$ = 792 nm periodic integrated SNSPDs.

In *P*-pitch integrated devices the polarization contrast more significantly depends on the profile of the wavelength-scaled nano-cavity-grating, and non-monotonously varies, when the polar angle is increased (Fig. 4b). A considerable polarization contrast modulation appears ~20.0° polar angle in S-orientation of all *P*-pitch integrated devices.

There is no significant difference in the course of polarization contrast achievable in P-orientation of NCAI-SNSPDs with different pitches, while presence of wavelength-scaled deflectors and embedded trenches causes a noticeable polarization contrast modulation also in this orientation. This indicates that wavelength-scaled deflector- and trench-arrays play important role in polarization contrast modulation as well.

In NCAI-SNSPD the polarization contrast is larger in S-orientation throughout the entire polar angle interval, except through the region corresponding to strong absorption modulation. At the center of this polar angle region the absorption of p-polarized light almost equals to the absorption of s-polarized light. The contrast exhibits a ($1.6 \cdot 10^1$) local maximum at 20.8° in S-orientation, while the largest $1.9 \cdot 10^2$ contrast is achieved at ~80.0° tilting. Slightly and significantly larger polarization contrast is attained throughout the entire polar angle interval in S-orientation of NCDAI- and NCDDAI-SNSPDs, respectively. Local $2.9 \cdot 10^2$ and $1.3 \cdot 10^5$ polarization contrast maxima appear at 18.2° and 20.6° angles, according to locally increased NbN absorption of p-polarized light in close proximity of these tilting in S-orientation of NCDAI- and NCDDAI-SNSPD. The contrast reaches $1.2 \cdot 10^3$ and $1.7 \cdot 10^5$ global maxima at 80.0° tilting, respectively. Considerably large $4.8 \cdot 10^4$ polarization contrast is achieved already at perpendicular incidence onto NCDDAI-SNSPD, which is important for practical applications. In S-orientation of NCTAI-SNSPD the course of polarization contrast is very similar to that in NCDDAI-SNSPD, however exhibits significantly smaller maximum of $5.3 \cdot 10^1$ at smaller 20.5° polar angle. Although, the global polarization contrast maximum is reached at small tilting in S-orientation, embedded trenches are on average detrimental on polarization contrast. In P-orientation of NCDAI- and NCTAI-SNSPDs the s-to-p contrast is smaller than unity at tilting corresponding to global minimum of polar angle dependent absorption in s-onto-P configuration (Fig. 3b).

Compared to *p*-pitch devices, the wavelength-scaled deflector-arrays result in significantly larger enhancement in the achievable polarization contrast, and the contrast enhancement is larger in presence in two deflectors in *P*-pitch integrated SNSPDs.

Although, the polarization contrast is slightly increased in S-orientation of *P*-pitch NCTAI-SNSPDs with respect to the contrast achievable in NCAI-SNSPD unlike counterpart *p*-pitch designs, embedded trenches are not capable of promoting selective read-out of quantum information encoded into polarization.

*3.2 Wavelength dependent NbN absorptance and polarization contrast*

Inspection of the spectral response in 500 nm wavelength interval [1250 nm, 1750 nm] around 1550 nm and in the entire polar angle interval, which can be investigated experimentally, was performed in p-onto-S and s-onto-P configurations of plasmonic structure integrated SNSPDs. The results proved that the sensitivity of these devices strongly depends both on their geometry and on the illumination conditions.

All integrated devices possessing a three-quarter-wavelength-scaled periodicity act as a frequency selective surface. As a result, a characteristic modulation is expected on the spectral response and on the dispersion diagram of *p*-pitch NCDAI-SNSPD, as well as of all *P*-pitch integrated SNSPDs (Fig. 5a, ab, Fig. 5b with insets).

*3.2.1 Wavelength dependent NbN absorptance: p-polarization*

Illumination by p-polarized light in S-orientation is capable of resulting in more characteristic modulation, since the intensity modulation is perpendicular to the periodic integrated pattern under these circumstances (Fig. 5-to-6). The spectral responses of *p*-pitch SNSPDs indicate that by increasing the wavelength, the NbN absorptance monotonously increases through the investigated spectral interval at any polar angle, except in NCTAI-SNSPD (Fig 5a with insets). The NbN absorptance maximum attainable at tilting corresponding to PBA increases with the wavelength in NCAI-SNSPD (Fig. 5aa). Larger slope modulation appears at larger wavelength, when the tilting of NCDAI-SNSPD is raised (Fig. 5ab). The achievable NbN absorptance monotonously increases by increasing the wavelength and by decreasing tilting of NCDDAI-SNSPD (Fig. 5ac). The NbN absorptance exhibits a maximum in wavelength at small tilting of NCTAI-SNSPD, and the attained maximum increases by decreasing tilting (Fig. 5ad).

At the maxima observed in polar angle the NbN absorptance slightly increases monotonously with the wavelength in S-orientation of all *p*-pitch integrated SNSPDs, except in NCTAI-SNSPD (Fig. 5a). The optical response of *p*-pitch NCAI- and NCDDAI-SNSPDs is the less sensitive to wavelength modification, according to their sub-wavelength period. In contrast, a characteristic modulation is observable exactly at 1550 nm on the spectral response of *p*-pitch NCDAI-SNSPD, which possesses a periodicity commensurate with $\lambda_{SPP, 1550 nm}$.

There is no modulation on the dispersion diagram of *p*-pitch NCAI-, NCDDAI- and NCTAI-SNSPDs according to their sub-wavelength period (Fig. 5ae, ag, ah). However, coupling on the deflector grating having a periodicity of $0.75*\lambda_{SPP, 1550 nm}$ in *p*-pitch NCDAI-SNSPD results in a modulation folded back into the first Brillouin zone (Fig. 5af).

The spectral response of *P*-pitch integrated devices' indicated that the NbN absorptance extrema including global minima in NCAI-SNSPD and global maxima in NCDAI/NCDDAI/NCTAI-SNSPDs appear at larger polar angle, when the wavelength is increased. The achieved NbN absorptances also increase with both variables (Fig. 5b with insets). The NbN absorptance indicates similar characteristic modulation starting at ~1400 nm in all SNSPDs consisting of 100 nm wide vertical gold segments separated by distances larger than 100 nm.

In *P*-pitch NCAI-SNSPD the NbN absorptance extrema appearing at small tilting shift rapidly forward in polar angle, when the wavelength is increased, while the NbN absorptance maxima appearing at PBA exhibit a weak wavelength dependency (Fig. 5ba) [19-21]. The inverted NbN absorptance extrema appearing in presence of deflectors and embedded trenches at small tilting exhibit significant wavelength sensitivity similarly to corresponding extrema in NCAI-SNSPD (Fig. 5ba-to-bb, bc, bd). Although, local NbN absorptance maxima appear already at perpendicular incidence onto these devices at ~1400 nm, the achieved NbN absorptance is smaller than at 1550 nm.

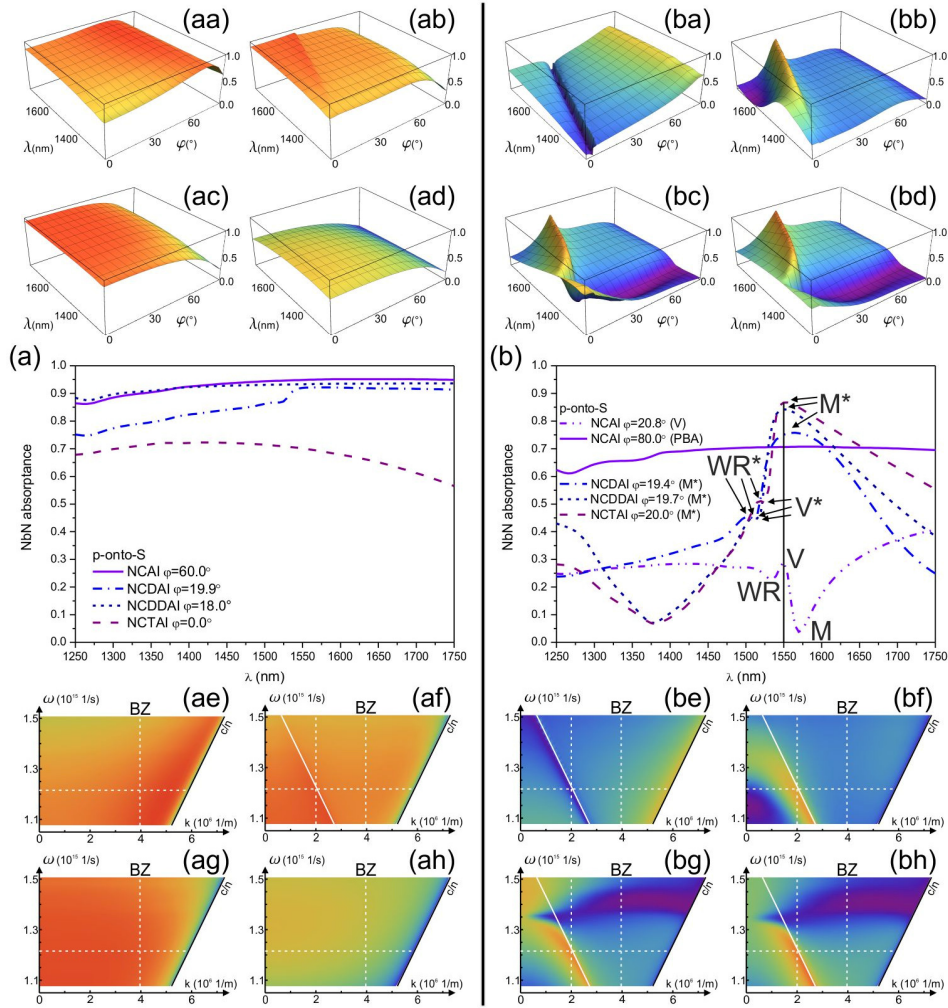

Fig. 5. NbN absorptance depicted as a function of wavelength and polar angle and the dispersion diagram in *p*-pitch and *P*-pitch (aa, ae and ba, be) NCAI-SNSPD, (ab, af and bb, bf) NCDAI-SNSPD, (ac, ag and bc, bg) NCDDAI-SNSPD, and (ad, ah and bd, bh) NCTAI-SNSPD illuminated by p-polarized light. Comparison of the wavelength dependent NbN absorptance of different (a) *p*-pitch and (b) *P*-pitch designs at the extrema in polar angle, which are of potential interest to practical applications. Those NbN absorptance extrema, which are of the same type in polar angle and wavelength, are indicated analogously on both graphs (M, V, WR, M*, V*, WR* points).

At the maxima observable during tilting in S-orientation of different *P*-pitch designs the NbN absorptance exhibits significantly different wavelength dependency (Fig. 5b). The NCAI-SNSPD spectrum computed for the local maximum (V point) in polar angle exhibits pseudo plasmonic band-gap (PBG) characteristics, since the global minimum in absorptance corresponds to global minimum/maximum in transmittance/reflectance (Fig. 5ba, be) [15, 22-4, 25-27]. At 80.0° tilting corresponding to PBA, the NbN absorptance remains constant through a wide wavelength region, which is advantageous for practical applications [19-21]. In NCDAI/NCDDAI/NCTAI-SNSPD designs all PBG features are completely inverted compared to the corresponding extrema in NCAI-SNSPD (Fig. 5bb-bd).

The global maxima appear inside an inverted pseudo PBG, namely in plasmonic pass-band (PPB) regions, since the global maximum in absorptance corresponds to global maximum/minimum in transmittance/reflectance (Fig. 5bf-bh). Moreover, coexistent global NbN absorptance minima are observable in presence of double deflectors and embedded trenches, which result in appearance of 492 nm wide empty cavities. However, the global NbN absorptance maximum appears at 1565 nm at 19.4° tilting (M* point) in NCDAI-SNSPD, i.e. it is slightly detuned from the desired wavelength, in NCDDAI- and NCTAI-SNSPDs the maximum appears exactly at 1550 nm. These NbN absorptance peaks override the values observable in NCAI-SNSPD at PBA through ~100 nm wavelength intervals. Both the maximum and FWHM of the wavelength-dependent NbN absorptance peaks is the largest in presence of embedded trenches, while double-deflectors result in a slightly smaller absorptance with an intermediate bandwidth. Complementary computations have shown that by increasing the incidence angle in NCAI-SNSPD, all PBG features shift forward, ensuring appearance of the same type extrema at 1550 nm as those observed on the polar angle dependent NbN absorptance, in accordance with the literature about Wood-anomalies [28-30].

The dispersion diagrams of all integrated *P*-pitch devices indicate well-defined modulations, which correspond to low energy plasmonic branches below the light line folded back into the first Brillouin zone (Fig. 5be-bh). In NCAI-SNSPD pseudo PBG appears at 1550 nm (Fig. 5be), while in presence of deflectors and embedded trenches curved PPBs appear in similar region revealing that propagating plasmonic modes are at play in NCDAI-, NCDDAI-, and NCTAI-SNSPDs (Fig. 5bf-bh). In NCDDAI- and NCTAI-SNSPDs a tilting independent lifted flat band corresponding resonance in empty 492 nm wide cavities intersects the PPB features, and manifests itself in global NbN absorptance minimum (Fig. 5bg, bh).

*3.2.2 Wavelength dependent NbN absorptance: s-polarization*

When s-polarized light illuminates *p*-pitch integrated NCAI-, NCDAI-, and NCDDAI-SNSPDs, the NbN absorptance increases by increasing the wavelength and by decreasing the angle of incidence at large wavelengths, while local NbN absorptance maxima are observable in polar angle at the small-wavelength-edge of the inspected interval (Fig. 6aa-ac). In NCTAI-SNSPD NbN absorptance maximum appears also in wavelength at small tilting, while the NbN absorptance maximum in polar angle, which is attained at large tilting, gradually increases towards the small-wavelength-edge (Fig. 6ad).

At the maxima in polar angle observable in s-onto-P configuration of NCAI-SNSPD the NbN absorption curve exhibits a similar characteristic as in p-onto-S configuration, however the achieved absorptance is significantly lower (Fig. 6aa). NCDAI-SNSPD ensures larger absorptance than NCAI-SNSPD in the entire investigated wavelength interval, while the course of wavelength dependent NbN absorptance is similar (Fig. 6ab). The NbN absorptance achievable in *p*-pitch NCDDAI-SNSPD is almost the same as in p-onto-S configuration and overrides the absorptances in NCDAI-SNSPD (Fig. 6ac). In NCTAI-SNSPD the NbN absorptance exhibits a maximum in wavelength, similarly to p-onto-S configuration, while the achieved maximal absorptance is significantly smaller than the maxima in other devices (Fig. 6ad). There is no characteristic modulation observable around 1550 nm on the spectral response of any *p*-pitch (Fig. 6a).

The dispersion diagram of *p*-pitch integrated devices does not indicate significant modulation, even in case of NCDAI-SNSPD possessing a periodicity of 792 nm (Fig. 6ae-ah).

The NbN absorptance exhibits a spectral dependency in s-onto-P configuration of different *P*-pitch designs, which is completely different and less significant than the sensitivity observed in p-onto-S configuration (Fig. 6b-to-5b). In NCAI-SNSPD the NbN absorptance maximum appearing at intermediate polar angles increases with the wavelength (Fig. 6ba). In NCDAI-SNSPD sudden NbN absorptance modifications occur in a wavelength and polar angle interval, which is complementary of the parameter-region corresponding to spectral modulation in case of p-polarized light illumination (Fig. 6bb-to-5bb). In NCDDAI- and NCTAI-SNSPD the NbN absorptance increases by decreasing either the wavelength or the polar angle (Fig. 6bc, bd).

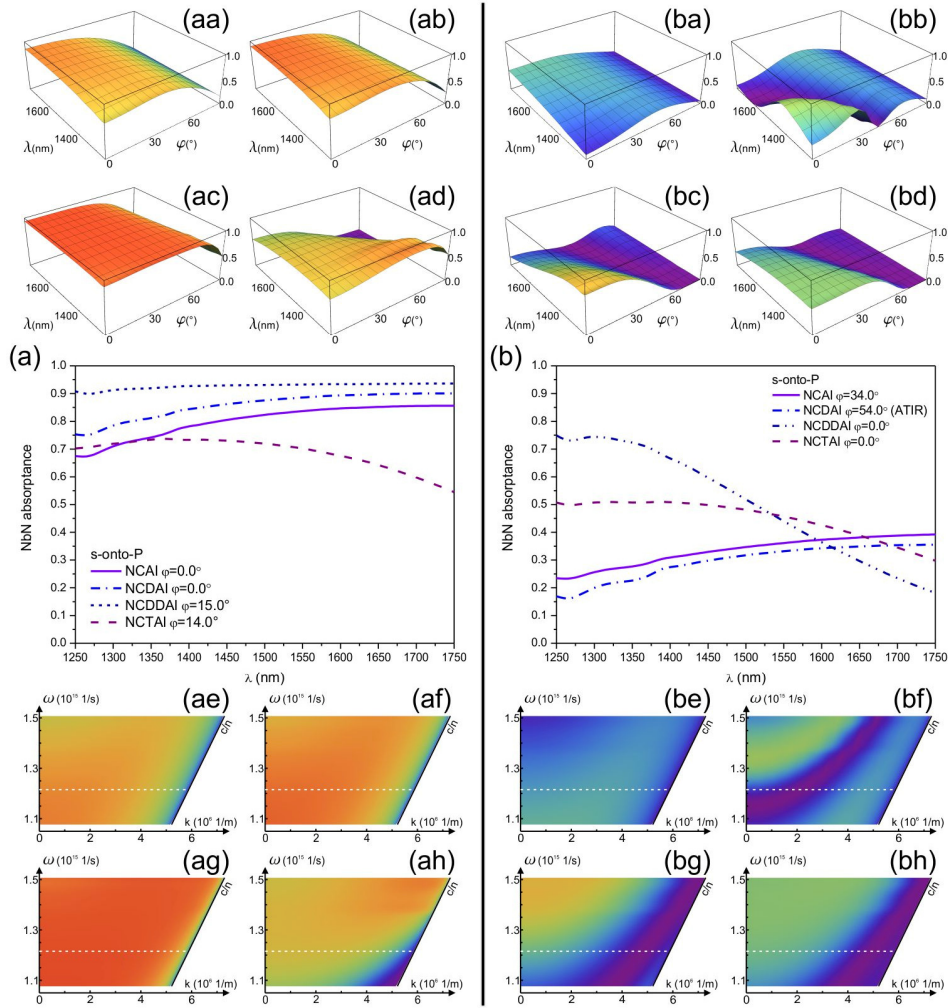

Fig. 6. NbN absorptance depicted as a function of wavelength and angle of incidence and dispersion diagram in *p*-pitch and *P*-pitch (aa, ae and ba, be) NCAI-SNSPD, (ab, af and bb, bf) NCDAI-SNSPD, (ac, ag and bc, bg) NCDDAI-SNSPD, and (ad, ah and bd, bh) NCTAI-SNSPD illuminated by s-polarized light. Comparison of the wavelength dependent NbN absorptance of different (a) *p*-pitch and (b) *P*-pitch designs at the extrema in polar angle, which are of potential interest to practical applications.

At maxima in polar angle in s-to-P configuration the NbN absorptance monotonously increases in the investigated wavelength interval both in NCAI- and NCDAI-SNSPDs, while in NCDDAI-SNSPD monotonous NbN absorptance decrease is observable from a considerably larger value and with a larger rate (Fig. 6b). In NCTAI-SNSPD the course of the NbN absorptance is similar to that in *p*-pitch design, however the achieved global NbN absorptance maximum is significantly smaller caused by three-times smaller fill-factor.

Dispersion diagrams for s-polarized illumination of *P*-pitch NCAI-, NCDAI-, NCTAI-, and NCDDAI-SNSPDs exhibit curved branches, which intersects the light line without local modulation. Although the p-onto-S and s-onto-P configurations are equivalent in the sense that coupled resonances can be excited due to **E**-field oscillation perpendicularly to the pattern, grating-coupling related branches are not observable in latter case.

Disappearance of coupled branches is caused by the perpendicularity of the $k_{photonic} \sin \varphi$ photonic wave vector projection to the $k_{grating}$ wave number, which proves that azimuthal orientation plays an important role in appearance of grating coupling related modulation. The slightly upward curved branches observable on s-polarized dispersion diagrams are in accordance with previous results in the literature on dispersion diagrams of short pitch gratings in equvivalent s-to-P configuration [22, 23].

*3.2.3 Wavelength dependent polarization contrast*

The spectral sensitivity of the polarization contrast was determined in 500 nm wavelength interval [1250 nm, 1750 nm] for all maxima appearing in polar angle in S- and P-orientations of plasmonic structure integrated SNSPDs (Fig. 7).

In *p*-pitch integrated devices the polarization contrast slightly increases monotonously by increasing the wavelength in all configurations (Fig. 7a). In NCAI-SNSPD the polarization contrast is noticeably larger at 60.0° polar angle in S-orientation because of the polarization selectivity of the PBA. There is no difference between polarization contrasts achieved at the global maxima in S- and P-orientation of NCDAI-, NCDDAI, and NCTAI-SNSDPs. The polarization contrast in S-orientation of NCAI-SNSPD is commensurate with the polarization contrast in either orientation of NCDAI-SNSPD, while with two orders of magnitude larger contrast is achieved in NCDDAI-SNSPD. This indicates that double deflectors have significant polarization contrast improving role in a wide bandwidth. The contrast values are the least of all in case of NCTAI-SNSPD.

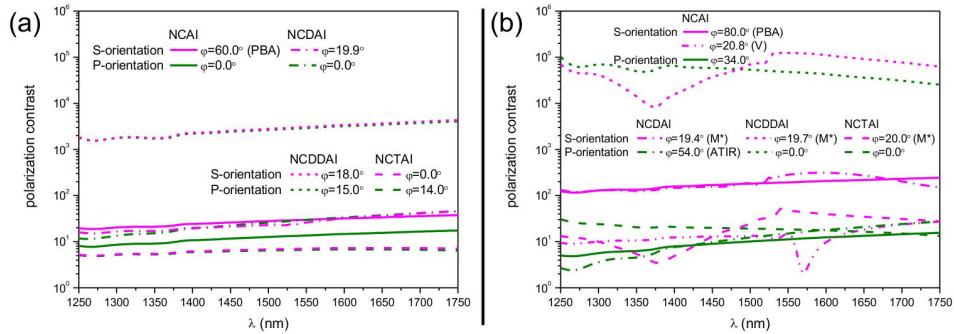

Fig. 7. Wavelength dependent polarization contrast at the maxima in polar angle in (purple) S-orientation and (olive) in P-orientation of (a) $p$ = 264 nm, (b) $P$ = 792 nm periodic integrated SNSPDs.

In *P*-pitch integrated devices the spectral sensitivity of the polarization contrast more significantly depends on the profile of the wavelength scaled nano-cavity-grating (Fig. 7b). The polarization contrast monotonously increases in both orientations of NCAI-SNSPD at the global maxima on polar angle dependent absorption. There is a considerable difference between them, because of the polarization selectivity of PBA in S-orientation. The contrast is non-monotonous at the V point in S-orientation, it clearly follows the course of wavelength dependent absorption. The wavelength dependent contrast at 19.4° polar angle in S-orientation of NCDAI-SNSPD increases monotonously and shows a clear modulation at 1550 nm wavelength, while at 54.0° polar angle in P-orientation the contrast is smaller with more than an order of a magnitude. The contrast reaches the highest values in presence of double deflectors. At 19.7° polar angle in S-orientation of NCDDAI-SNSPD the contrast curve indicate features originating from the plasmonic pass-band and flat-band regions, while at perpendicular incidence in P-orientation the contrast decreases almost monotonously. The course of polarization contrast is similar in counterpart S- and P-orientation of NCTAI-SNSPD at 20.0° and 0.0° tilting, however the achieved contrast values are three orders of magnitude lower than in the presence of double deflectors.

In comparison, commensurate polarization contrast is achieved in P-orientation of *p*- and *P*-pitch NCAI-SNSPDs, while one order of magnitude larger polarization contrast is observable in S-orientation of all *P*-pitch devices. The polarization contrast attainable in NCDAI-SNSPD is commensurate with the contrast in analogous configurations of NCAI-SNSPD, while two and three orders of magnitude larger contrast is achievable in *p*- and *P*-pitch NCDDAI-SNSPDs, respectively. These results prove that wavelength-scaled double-deflector-arrays can result in polarization contrast enhancement throughout wide spectral intervals.

*3.3 Time-averaged and time-dependent near-field*

A detailed near-field study was performed to uncover the physical origin of optical response modulation in different SNSPDs.

*3.3.1 Time-averaged and time-dependent near-field: p-polarized illumination*

In *p*-pitch integrated SNSPD devices at tilting corresponding to global NbN absorptance maxima the normalized **E**-field is enhanced at the entrance of all MIM cavities, however the origin of this enhancement, and the distribution along the nano-cavity array depends on the integrated grating profile. In *p*-pitch NCAI-SNSPD (Fig. 8aa, Media 1) the uniform **E**-field enhancement is due to tunneling of forward propagating waves through all MIM nano-cavities at the 60.0° polar angle corresponding to PBA. In *p*-pitch NCDAI-SNSPD the normalized **E**-field is most strongly enhanced in those MIM nano-cavities, which are in the closest proximity of deflectors (Fig. 8ab, Media 2). The plasmon-wavelength-scaled deflector array causes the highest **E**-field enhancement at 19.9° polar angle corresponding to PPB center in counterpart *P*-pitch device. Although, the normalized **E**-field is enhanced slightly in all MIM nano-cavities due to collective resonances on the sub-wavelength array, well directed power-flow is observable exclusively in the nano-cavities preceding deflectors. In addition to this, backward propagating surface waves are observable below the boundary pair (Media 1). In *p*-pitch NCDDAI-SNSPD the normalized **E**-field is again uniformly enhanced in all MIM nano-cavities at the global maximum appearing at 18.0° due to efficient collective resonances on the sub-wavelength array, and the enhancement is larger than in NCAI-SNSPD at similar small tilting, since the power-flow is enhanced by waves gradually squeezed and quided by double deflectors (Fig. 8ac, Media 3). In *p*-pitch NCTAI-SNSPD enhanced **E**-field and strong power-flow is observable alternately at the entrance of empty and filled MIM nano-cavities at perpendicular incidence, respectively. However, the **E**-field enhancement is larger at the entrance of empty embedded trenches, hence the achievable NbN absorption is significantly smaller than in other *p*-pitch SNSPD in contempt of larger powerflow towards NbN segments (Fig. 8ad, Media 4).

In *P*-pitch NCAI-SNSPD at the 19.7° global minimum neither the normalized **E**-field maxima nor the counterclockwise power-flow vortices approaches the MIM nano-cavities but are concentrated under the vertical gold segments center (Fig. 8baa, M point, Media 5), as a consequence low absorptance is attainable. At the 20.8° local maximum both the normalized **E**-field maxima and the clockwise (below cavity) and counterclockwise (below gold segment) power-vortices are shifted laterally, as a result enhanced power-flow is noticeable towards the NbN segments at the entrance of MIM nano-cavities (Fig. 8bab, V point, Media 6). The left-side down oriented net power out-flow and the time evolution of the $\sqrt{E_x^2 + E_y^2 + E_z^2}$ quantity indicate that the composite scattered field comprises backward propagating weakly bounded surface waves as well as reflected waves at these extrema (Fig. 8baa-bab, Media 5, 6). At the M point the surface waves exhibit more bounded characteristics, while below the boundary pair intense specular back-reflected waves are observable. The wavelength of the surface waves at the M point equals to the SPP's wavelength, while at the V point it is slightly larger than the wavelength in silica, indicating that they are Brewster-Zenneck-type waves [28-30].

At the 80.0° plasmonic Brewster-angle the power-flow is directed forward almost parallel to the interface according to the forward propagating waves observable with wave fronts almost perpendicular to the silica interface, which illuminate the neighboring MIM cavities intermittently (Fig. 8bac, PBA point, Media 7). Their transversal wave vector equals to the wave vector of Brewster-Zenneck waves at V point indicating that the backward and forward propagating waves, which play important role in absorptance enhancement at local and global maxima, have similar characteristics. In contempt of wave vector matching, the absorptance is smaller than in counterpart *p*-pitch design, caused by three-times smaller fill-factor.

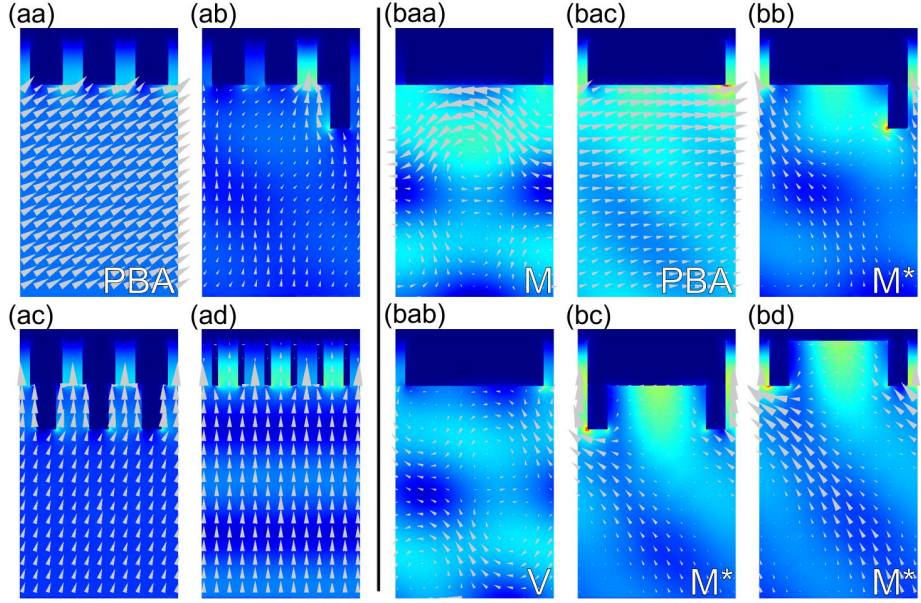

Fig. 8. Time-averaged **E**-field with power-flow arrows in p-onto-S configuration of *p*-pitch and *P*-pitch NCAI-SNSPD (aa) 60.0° (PBA point), (baa) 19.7° (M point), (bab) 20.8° (V point) and (bac) 80.0° (PBA point); NCDAI-SNSPD (ab) 19.9° and (bb) 19.4° (M* point); NCDDAI-SNSPD (ac) 18.0° and (bc) 19.7° (M* point); NCTAI-SNSPD (ad) 0.0° and (bd) 20.0° (M* point). Time evolution of the **E**-field ($\sqrt{E_x^2 + E_y^2 + E_z^2}$ quantity) is provided in corresponding multimedia files 1-10 (enhanced online).

Fig. 8bb, bc, bd indicate that at the 19.4°, 19.7° and 20.0° global absorptance maxima (M* points) in *P*-pitch NCDAI-, NCDDAI- and NCTAI-SNSPD the hot-spots of normalized **E**-field coincide with the NbN segments. Strongly enhanced power-flow is directed into all nano-cavities neighboring 100 nm wide gold segments acting as deflectors or walls of embedded trenches. The corresponding multimedia files show weakly bounded surface waves propagating backward below the nano-cavity array (Media 8-10). Although, an **E**-field enhancement occurs at the entrance of neighboring cavities alternately, the NbN segments are shined very efficiently. As a result of perfect synchronization via three-quarter wavelength periodicity, the backward propagating waves promote the illumination of NbN segments at the entrance of quarter wavelength nano-cavities. The backward propagating waves have a wavelength equal to $\lambda_{SPP,\ 1550\ nm}$, however their decay length is larger than $\delta_{SPP,\ 1550\ nm}$ indicating that the modes resulting in **E**-field enhancement in presence of inserted and embedded deflectors differ from SPPs. Clockwise power-flow vortices are observable below the gold segments in NCDAI-SNSPD (Fig. 8bb). In NCDDAI- and NCTAI-SNSPDs significant **E**-field enhancement is observable between double deflectors and inside embedded trenches as well.

The power-flow is vortices-free and counterclockwise oriented parallel to the substrate interface in NCDDAI- and NCTAI-SNSPDs (Fig. 8bc and bd), proving the advantages of symmetric profiles. The similarity of power-flows indicates that the vertical gold segments in NCTAI-SNSPD can be considered as the most simple embedded deflector array.

*3.3.2 Time-averaged and time-dependent near-field: s-polarized illumination*

In equivalent s-to-P configuration the normalized **E**-field is enhanced in all MIM nano-cavities due to collective resonances on their sub-wavelength array. In *p*-pitch NCAI- and NCDDAI-SNSPD the normalized **E**-field is uniformly enhanced in all MIM nano-cavities at the 0.0° and 15.0° global maximum, respectively (Fig. 9aa, Media 11 and 9ac, Media 13). Comparison of **E**-field time-dependencies shows that the NbN segments are more efficiently illuminated in NCDDAI-SNSPD in contempt of local enhancements at deflector corners, due to guidance of gradually squeezed modes between deflectors towards the nano-cavities.

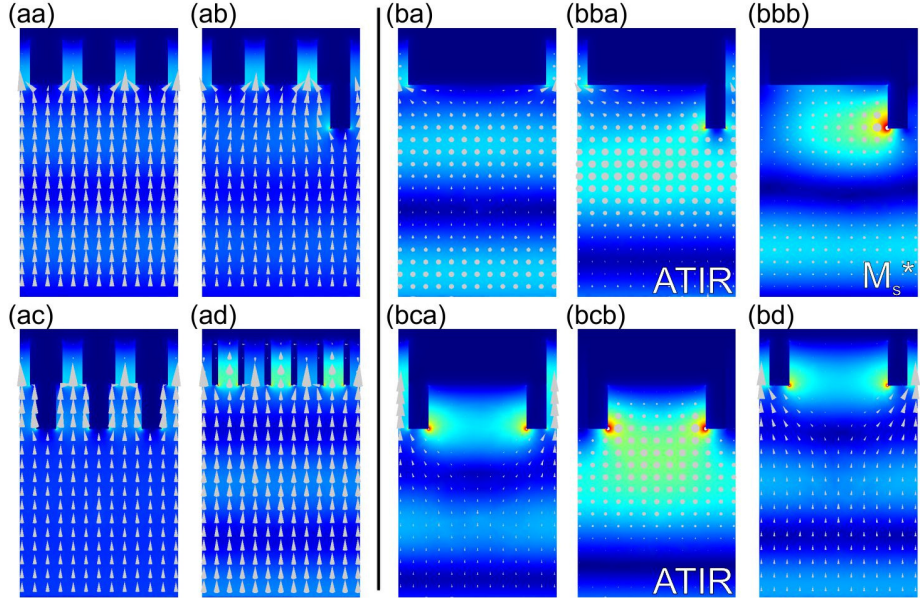

Fig. 9. Time-averaged **E**-field with power-flow arrows in s-onto-P configuration of *p*-pitch and *P*-pitch NCAI-SNSPD (aa) 0.0° and (ba) 34.0°; NCDAI-SNSPD (ab) at 0.0° and (bba) 54.0° (TIR) and (bbb) 21.0° ($M_s^*$), NCDDAI-SNSPD (ac) 15.0° and (bca) 0.0° and (bcb) 56.0° (TIR), NCTAI-SNSPD (ad) 14.0° and (bd) 0.0°. The time evolution of the **E**-field ($\sqrt{E_x^2 + E_y^2 + E_z^2}$ quantity) is provided in corresponding multimedia files 11-20 (enhanced online).

In *p*-pitch NCAI-SNSPD the normalized **E**-field is most strongly enhanced again in MIM nano-cavities in the closest proximity of deflectors at perpendicular incidence corresponding to global maximum. However, the **E**-field and power-flow enhancement with respect to other cavities is smaller than at the global maximum in p-onto-S configuration, and below the boudary pair there are no backward propagating waves (Fig. 9ab, Media 12). In *p*-pitch NCTAI-SNSPD, although the normalized **E**-field is uniformly enhanced in all MIM nano-cavities at the 14.0° global maximum (Fig. 9ad, Media 14), the enhancement is larger inside embedded trenches. The Poynting-vector exhibits a complement characteristics, namely the power-flow is larger through the NbN contaning cavities, similarly to p-onto-S configuration.

In *P*-pitch NCAI-SNSPD the normalized **E**-field is uniformly enhanced in all MIM nano-cavities at the 34.0° global maximum, but the attained absorption is smaller than in *p*-pitch design because of three times smaller fill-factor (Fig. 9ba, Media 15).

In *P*-pitch NCDAI-SNSPD at the 54.0° global maximum the normalized **E**-field is alternately enhanced at the entrance of the MIM nano-cavities neighboring the deflectors on their leading and leaving side (Fig. 9bba, ATIR point, Media 16). A weak sideward power-flow is noticeable towards the cavities preceding the deflectors in P-orientation as well. In addition to this, the phase shift accompanying the TIR phenomenon in case of s-polarized light illumination is capable of contributing to the **E**-field enhancement at the entrance of the MIM nano-cavities at this large tilting [31]. At tilting corresponding to global NbN maximum in p-to-S configuration no surface waves are excited by s-polarized illumination, on the contrary, the **E**-field is trapped at the top of deflectors causing a global minimum (Fig. 9bbb, M*s point, Media 17). In *P*-pitch NCDDAI- and NCTAI-SNSPD at the global maxima appearing at perpendicular incidence the normalized **E**-field is uniformly enhanced and strong power-flow is observable into all MIM nano-cavities (Fig. 9bca, Movie 18 and Fig. 9bd, Movie 20). Larger NbN absorptance is attainable in NCTAI-SNSPD, since the **E**-field enhancement occurs in closer proximity of the absorbing NbN segments in presence of embedded deflectors. During tilting significant part of the **E**-field is concentrated inside the 492 nm wide empty-cavity-like regions mainly at the top of deflectors. This **E**-field distribution limits the achievable enhancement in the NbN conaining nano-cavities, and results in global minimum in both designs close to the global maximum in NCDAI-SNSPD (Fig. 9bcb, ATIR point, Movie 19).

## 4. Discussion

Our present results prove that unique nanophotonical phenomena occur in plasmonic structure integrated SNSPDs, which result in significant optical response and near-field modulation. Geometrical and optical properties of the integrated detectors together determine the illumination direction and wavelength region, where these unique nanophotonical phenomena are capable of enhancing NbN absorptance. In equivalent p-onto-S and s-onto-P configurations of plasmonic structure integrated devices the **E**-field oscillation perpendicular to the vertical noble-metal segments ensures excitation of resonant localized plasmonic modes in the NbN containing MIM nano-cavities (Fig. 2, 3, 5, 6, 8, 9). Appropriate tilting in S-orientation ensures the most efficient excitation of coupled resonances (Fig. 2 - to- Fig. 3).

Both p- and P-pitch integrated MIM patterns act as loaded nano-cavity gratings, where positioning of the NbN segments at the entrance of quarter wavelength cavities ensures overlap with high intensity EM-field. The strongly sub-wavelength periodicity of *p*-pitch designs makes possible almost polar angle independent collective resonances both in p-onto-S and s-onto-P configurations. The NbN absorptance is perturbed by p-polarization specific phenomena, namely via plasmonic Brewster angle (Fig. 2aa, 5aa, 5ae, 8aa) in NCAI-SNSPDs and via plasmonic pass-band features in NCDAI-SNSPD (Fig. 2ab, 5ab, 5af, 8ab). P-polarized light illumination of integrated systems possessing *P*-periodicity commensurate with $0.75*\lambda_{SPP,\ 1550\ nm}$ results in coupled resonances with tilting dependent efficiency and in appearance of pseudo PBG and PBA features in NCAI-SNSPD (Fig. 2, 5b, 5ba), and plasmonic pass-bands in NCDAI-, NCDDAI- and NCTAI-SNSPD (Fig. 2, 5b, 5bb-d) [15, 22-24, 25-27]. The most advantageous p-onto-S configurations are, where specific series of MIM cavities are synchronously illuminated, which promotes collective resonance phenomena [14, 15, 22-24]. All presented PBG and PPB features appear in S-orientation of *P*-pitch devices around orientation corresponding to synchronous illumination of series made of four cavities in the integrated patterns:

$$\sin \varphi_{m,k} = (m\lambda_{photonic})/(kP), \qquad (1)$$

i.e. when *m*=1 and *k*=4. Fingerprint of PBG in *p*-pitch NCDAI-SNSPD is observable at polar angle, where *k*=12, in accordance with the smaller period. The condition of Wood-anomaly, which results in coupling to photonic modes on $k*P$ periodic grating, corresponds to polar angles between the global extrema (M/M* points) and the neighboring local extrema (V/V* points) (Fig. 2b, 5b, 8ab, 8baa (M point) and 8bab (V point), bb-bd (M* points)) [28-30].

The observed pseudo PBG and PPB features results in typical Fano-spectra, which originate from coupling between resonant localized and non-resonant surface modes based on the literature [35-38]. The inspection of the near-field proved the co-existence of different surface modes, in accordance with composite diffraction evanescent waves (CDEW) theory [39-45]. The wavelength of observed surface modes, which are grating-coupled at the NbN absorptance extrema in p-to-S configuration of $P$-pitch SNSPDs can be computed as:

$$-k_{surface\ wave} = (k_{photonic}\sin\varphi) - k_{grating}. \qquad (2)$$

Accordingly, the modulation on the dispersion diagrams reveals that the $p$-pitch NCDAI-SNSPD as well as all $P$-pitch integrated patterns act as a deep short pitch gratings, which results in first order coupling (Fig. 5ab, ba-bd) [22-24]. The grating-coupled modes at the global extrema have uniformly a wavelength equal to $\lambda_{SPP}$ (Fig. 2b, 5b, 8ab and Media 2, 8baa and Media 5 (M point) and 8bb-bd and Media 8-10 (M* points)). All local extrema (Fig. 2b, 5ab, 5b, Fig. 8 bab and Media 6 (V point)) appear at orientations, where grating-coupling dominantly results in waves having a wavelength longer than $\lambda_{SPP}$ and a transversal decay length larger than $\delta_{SPP}$, similarly to the initial non-bounded waves described in CDEW theory [43]. These modes appear in case of non-metallic bounding media as well, proving that they are analogous with Brewster-Zenneck waves [14, 15, 31-33].

However, the angle dependent optical responses also strongly depend on the grating profile. In NCAI-SNSPD the appearance of the global minimum is caused by de-synchronization of SPPs with respect to the MIM cavities (Fig. 8baa, M point) [14, 15, 25-27]. According to this, the SPPs are subsequently coupled to propagating modes at the M point, as predicted by dynamical diffraction theory of gratings [39], rather than promote absorption in NbN segments (Fig. 8baa, Media 5).

The global maximum in $p$- and $P$-pitch NCAI-SNSPDs appears at the 60.0° and 80.0° plasmonic Brewster-angle (Fig. 2a, b, 8aa, bac PBA points) specified by

$$\cos\varphi = (k_{surface\ wave}w)/(k_{photonic}p), \qquad (3)$$

where $w$ refers to the MIM cavity's width. At PBA transversal wave-number component of the incoming light equals to the wave-number of Brewster-Zenneck modes that are observable at the local extrema in $P$-pitch NCDAI/NCTAI/NCDDAI-SNSPDs. The appearance of the global NbN absorptance maximum is explained by coupling to surface waves capable of illuminating the MIM nano-cavities, and by subsequent efficient tunneling of light through them [19-21]. However, light in-coupling at this large tilting might be realized exclusively via prism, or via appropriately cut fibers.

The integrated deflectors, double-deflectors and embedded trenches with a periodicity of $0.75*\lambda_{SPP}$ in $P$-pitch NCDAI/NCTAI/NCDDAI-SNSPDs act as array of plasmonic mirrors with different efficiency, which realize conversion of the modes incident at grazing angle, and ensure their phase matching at the entrance of MIM nano-cavities [18, 46-48]. In addition to this, the deflectors compose a lateral cavity array with $1.5*\lambda_{SPP}$ characteristic length, which also promotes in phase reemission towards localized modes in MIM nano-cavities [46]. The inversion of the pseudo PBG and the appearance of global maximum at the PPB center is due to that the surface waves having a wavelength equal to $\lambda_{SPP}$ are converted into Brewster-Zenneck modes, back-reflected with good efficiency and laterally synchronized on the three-quarter wavelength periodic MIM nano-cavity-array (Fig. 5be-bh). In $P$-pitch devices the **E**-field oscillation direction promotes collective resonances in s-to-P configuration, however the intensity modulation at oblique incidence is parallel to the periodic integrated pattern (Fig. 6be-bh). As a consequence the grating coupling phenomena are not at play, and the attainable NbN absorptance is determined mainly by the profile of the pattern.

The asymmetric profile of NCDAI-SNSPD results in a global minimum instead of a maximum in absence p-polarization specific waves (Fig. 6bbb). The phase-shifts accompanying s-polarized light illumination in ATIR region and the sideward power-flow result in global maximum in presence of single deflectors, while double deflectors and embedded trenches trap the **E**-field in between NbN containing cavities (Fig. 6bba-to-bcb).

The spectral sensitivity of integrated devices under illumination conditions corresponding to NbN absorptance maxima in p-to-S and s-to-P configurations strongly depends on the grating profiles. The NbN absorptance is almost wavelength independent in *p*-pitch devices, as well as at the PBA in *P*-pitch NCAI-SNSPD, which is very important for applications. The pseudo PBG and PPB features appearing in p-to-S configuration of *P*-pitch integrated devices prove that they act as frequency selective surfaces. The highest degrees of freedom in spectral engineering and the largest number of nanophotonical phenomena capable of enhancing NbN absorption exist in p-to-S configuration. The peculiarity of the optimal configurations of the integrated systems is that these are neither at perpendicular incidence nor at the Brillouin zone boundaries. In contrast, these configurations correspond to $(k_{photonic} \sin \varphi)/k_{grating} = 0.5$ case, which results in same **E**-field in neighbouring nano-cavities [22-24].

The polarization sensitivity of the integrated systems shows that double deflectors play an important role in polarization contrast enhancement, while embedded trenches are detrimental in applications based on contrast maximization.

## 5. Conclusion

The presented results uncover general nanophotonical phenomena in different integrated device types, which result in analogous optical response and near-field modulation.

Important conclusion of our work is that each plasmonic structure integrated SNSPD design has its own unique optimal configuration. In case of presented integrated SNSPDs on silica substrate large absorptance is attainable at specific tilting in S-orientation with large bandwidth and with involvement of specific surface modes under specific conditions. Illumination of *p*-pitch integrated devices results in weakly polar angle and wavelength dependent NbN absorptance. This is due to that in all sub-wavelength periodic integrated SNSPDs the collective resonances on nano-cavity-arrays result in enhanced **E**-field, while in *P*-pitch integrated devices the tilting has significant impact on the achievable absorptance. In *p*/*P*-pitch NCAI-SNSPDs 95.3/70.3 % absorptance is attained at the plasmonic Brewster-angle, while 92.7/75.0 % - 93.3/84.3 % - 70.1/86.7 % absorptances are achieved at small tilting/the center of plasmonic-band-pass regions in NCDAI-, NCDDAI-, NCTAI-SNSPD devices. Present work is the first demonstration of 86.7 % NbN absorptance in wavelength-scaled (792 nm) periodic patterns, corresponding to a figure of merit (FOM ~ device efficiency/reset time), which is competitive with FOM previously attained exclusively via sub-wavelength patterns of wider superconducting wires. Moreover the spectral sensitivity of the integrated patterns makes possible to achieve ~90.0 % NbN absorptance through a ~100 nm wavelength interval. At tilting resulting in maximal absorptance the polarization contrast is commensurate in S- and P-orientation of *p*-pitch NCDAI- and NCDDAI-SNSPDs, while larger polarization contrast achieved in S-orientation of all other devices through a wide band surrounding the 1550 nm. The highest ~$10^4/10^5$ polarization contrast is reached in *p*/*P*-pitch NCDDAI-SNSPD, which open novel avenues in QIP applications.

Determination of absolutely optimal configurations of the inspected device designs based on their dispersion characteristics is in progress.

**Acknowledgments**

The authors would like to thank Francesco Marsilli, Sae Woo Nam and Áron Sipos for the helpful discussions. This work was partially supported by the European Union and the European Social Fund through project "Supercomputer, the national virtual lab" (grant no.: TÁMOP-4.2.2.C-11/1/KONV-2012-0010) and "Impulse lasers for use in materials science and biophotonics" (grant no.: TÁMOP-4.2.2.A-11/1/KONV-2012-0060).